\documentclass[aps,prl,preprint,superscriptaddress]{revtex4}
\usepackage{graphicx}
\usepackage{dcolumn}
\usepackage{bm}
\usepackage{epstopdf}
\usepackage{csquotes}
\usepackage{amsmath}
\usepackage{graphicx}
\usepackage{siunitx}
\usepackage[]{changes} 
\newcommand{\add}{\textcolor{black}}
\newcommand{\remove}[1]{}

\begin{document}


\title{Classical State Detection Using Quantum State Tomography}

\author{Kim Fook Lee and Prem Kumar}
\affiliation{%
Center for Photonic Communication and Computing, Department of Electrical Engineering and Computer Science, Northwestern University, 2145 Sheridan Road, Evanston, IL 60208-3112, USA}%
\email[]{kim.lee@northwestern.edu}



\date{\today}

\begin{abstract}
We present a model to detect a classical state mixed with an idler photon from a polarization-entangled pair. A weak coherent light with a well-defined polarization, matched in wavelength to the idler photon, is injected into the idler channel. Quantum state tomography is then performed on both the classically mixed idler photon and its entangled signal partner. The reconstructed state is modeled as a combination of an $X-$quantum state and a classical$-$quantum (CQ) state. In this framework, the weak coherent light acts as a measurement apparatus performing a local polarization measurement on the idler channel, thereby inducing a classical state. The density matrix of the classical state is identified via algorithmic analysis of the diagonal and off-diagonal elements of the reconstructed density matrix.
This approach could advance techniques for classical$-$quantum coexistence in networking applications$\,-\,$such as quantum wrapping$\,-\,$as well as future quantum key distribution protocols based on the coexistence of weak coherent states and entangled photon states.
\end{abstract}

\pacs{03.67.Hk, 42.50.Lc, 03.67.Mn, 42.50.Ar}

\maketitle


\section{1. Introduction}

A quantum channel\cite{LiNan_pra_2022,Modi_2010,Luo_2024} describes the flow of quantum information between two parties who share a bipartite quantum state, when a local operation and classical communication (LOCC) protocol\cite{LOCC_2009} is applied to one of the subsystems. Consider two parties, Alice and Bob, who share a quantum state $\rho_{AB}$. Suppose Alice performs a local quantum measurement on her subsystem, such that her superposition state $(|H\rangle + |V\rangle)/\sqrt{2}$ collapses to either $|H\rangle$ or $|V\rangle$. She then sends the classical information (either $H$ or $V$) to Bob via a classical communication channel. Based on this information, Bob performs a conditional quantum measurement on his subsystem and returns the result to Alice through the same classical channel. This process constitutes a standard LOCC protocol. In such processes, a measurement on subsystem A alters our knowledge of subsystem B. The flow of information$\,-\,$whether through quantum or classical channels$\,-\,$serves as a mechanism for extracting or inducing correlations within the bipartite system.

The extent to which quantum information in party B is modified by a local measurement on party A depends on the nature of the measurement, such as a local von Neumann projection. Several quantum information measures\cite{Modi_RMP_2012,Bera_2017}$\,-\,$such as quantum discord\cite{Zurek_prl_2001,Vedral_2001,Caves_prl_2008,Singh_2014,Luo_pra_2008,Dakic_2010,Fu_2010,Piani_2012,Chang_2013}, measurement-induced disturbance\cite{Luo_2008,LuoN_2011,Adesso_2011,Maziero_2009}, and measurement-induced entanglement\cite{Google_2023,LuoFu_2011,Xi_2012,Fan_2015,Li_2016,RR_2017}$\,-\,$have been introduced to characterize information flow within the framework of quantum and classical channels. We have recently employed quantum discord to investigate quantum correlations in photon pairs generated by green fluorescent protein\cite{Kflee_nc_2017,Shi_sr_2016}, as well as to study non-Markovian dynamics in a bipartite quantum system stored in a fiber-based quantum buffer\cite{kflee24,kflee25}.

Quantum discord\cite{Zurek_prl_2001,Vedral_2001} is a measure of quantum correlations, defined as the difference in total correlations of a bipartite state $\rho_{AB}$ before and after a local measurement is performed on one of the parties. It was introduced to account for environment-induced decoherence, where the environment is modeled as a measurement apparatus, or equivalently, as one party (e.g., Alice) performing a local measurement. In this context, the environment effectively measures Alice's photon, initially in a superposition state, thereby inducing a classical state in her channel. This classical component can be detected through coincidence and accidental-coincidence measurements performed on the bipartite state. We believe that the contribution of this classical component in $\rho_{AB}$ is relatively small and remains largely unexplored by standard quantum information processing techniques such as quantum state tomography\cite{James_pra_2001}.

Recently, classical-quantum data frames\cite{Shabani_22,BenYoo_2024,Gamze_2025} have been introduced in quantum networks to enable functions such as packet synchronization, label processing, and contention management. In these data frames, a coherent light source such as laser light with a well-defined polarization state coexists with quantum light in the same optical fiber, although the two operate at different optical wavelengths. The laser light can generate Raman scattering or other noise sources at wavelengths overlapping with the quantum signal. In addition, the coherent light source can act as a measurement apparatus, thereby inducing classical state mixing alongside the quantum light. When a quantum system is coupled to a classical state, a mixture of classical and quantum correlations can emerge. This is analogous to a von Neumann quantum measurement, where correlations are formed between a quantum system and a classical state as a result of the measurement process, leading to measurement-induced correlations\cite{Khalid_2018,Khalid_2020,Fu_2006,Illu_2006,SG_2012,illu_2013,Yu_2013,Berr_2012}.

In this paper, we inject a weak coherent light with a well-defined polarization state into a idler photon channel, which is generated from a fiber-based entangled photon source. We then perform a full quantum state tomography on the classically mixed idler photon and its paired signal photon. The weak coherent light is interpreted as a measurement apparatus, which introduces noise photons and induces a classical state at the same optical wavelength as the idler photon. This classical state corresponds to the polarization state of the injected weak coherent light. To extract information about the classical state, we assume the initial state is a Werner state\cite{Werner_pra_1989,Munro_pra_2001} and the final state is a $X-$ state\cite{Donnert_nmeth_2007,Luo_pra_2008}.
We model the final state as a combination of its quantum component and a classical$-$quantum (CQ) state. The CQ state arises from a local quantum operation (measurement) on the idler photon.
We numerically extract the density matrix of the classical state from the CQ state by solving the diagonal and off-diagonal elements of the measured final state's density matrix with the theoretical model. The resulting classical state's density matrix corresponds to the local quantum operation acting on the idler photon. The detected classical state agrees with the experimental result. We then evaluate the diagonal quantum discord of the states involved in this study. Our experimental approach demonstrates both the practicality and robustness of detecting classical state by using quantum state tomography.
This method could significantly influence the development of tools for classical-quantum coexistence in networking techniques. Furthermore,  our algorithm may prove highly beneficial for advancing quantum key distribution protocols that involve the coexistence of the weak coherent states\cite{Barbosa_2003,Yuen_2003,Decoy_2005,Sua_2011} and entangled photon states\cite{BB_2014,BB_1992,Gisin_2002}.

\section{2. Results}

The experimental setup is shown in Fig.1. We use a counter-propagating scheme (CPS) \cite{Kflee_ol_2006,Sua_ol_2014} to generate a two-photon polarization entangled state, $|\psi_{\circ}\rangle=\frac{1}{\sqrt{2}}[|H_s H_i\rangle + |V_s V_i\rangle]$, through a four-wave mixing process in 300$\,$m of dispersion-shifted fiber.
The signal and idler photons are separated by dense wavelength-division multiplexers (DWDMs). The signal (idler) photon is sent to the polarization analyzer $\rm{PA_{s}}$($\rm{PA_{i}}$), respectively. The first pair of a quarter-wave plate (QWP) and a half-wave plate (HWP) is used to align the photon polarization (H or V) along the principal axes of the second pair of QWP and HWP, and the polarization beam splitter (PBS). The second pair of QWP and HWP is used to perform quantum state tomography.
The idler photon is mixed with a weak coherent light field using a DWDM at the idler wavelength. This weak coherent light serves as the classical light in our experiment, and its polarization state represents a classical state. The resulting classically mixed idler photon is sent to the polarization analyzer $\rm{PA_{i}}$.
The classical light is obtained from a continuous-wave (CW) tunable fiber laser.
We use free-space optics to prepare its polarization state.
QWP1 and HWP1 are used to perform a task similar to that of the first pair of the QWP and HWP.
We then independently insert the HWP0 or the QWP0 or both to project a classical state such as $H$, $A=\frac{1}{\sqrt{2}}(H - V)$, and $L = \frac{1}{\sqrt{2}}(H - i\,V)$ in front of the $\rm{PA_{i}}$.

\begin{figure}
 \centering
        \includegraphics[width=0.7\textwidth]{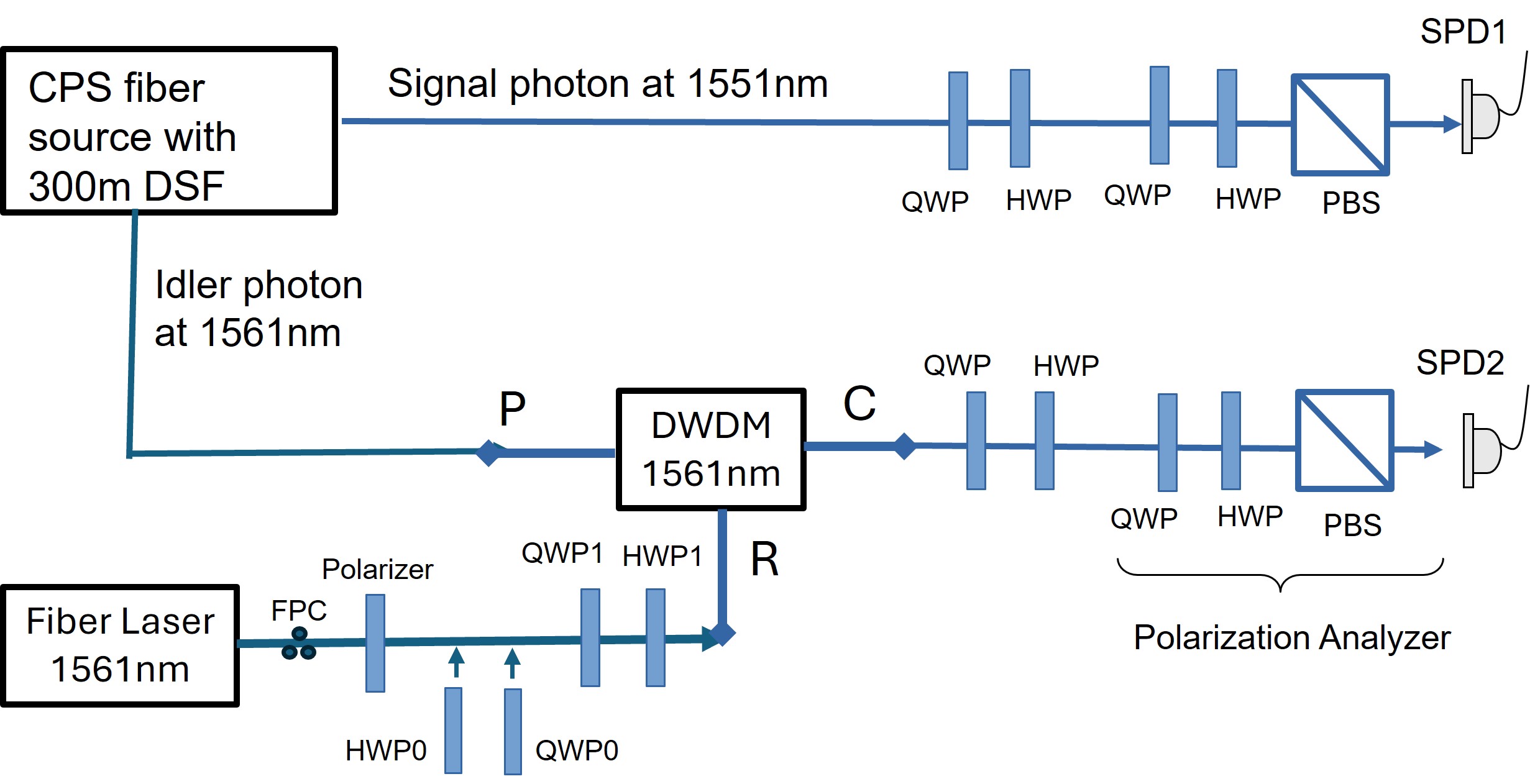}
 \caption{A counter-propagating scheme is used to generate two-photon polarization entangled photon pairs. The polarization state of the weak coherent light is prepared in free-space and then mixed with the idler photon through a DWDM. Transmission from port R to port C of the DWDM provides 23 dB of attenuation for weak coherent light at 1561 nm. The classically mixed idler photon and the signal photon are analyzed by two spatially separate polarization analyzers ($\rm{PA_{s,i}}$). HWP; Half-wave plate. QWP; Quarter-wave plate. PBS; polarization beam splitter. SPD; Single photon detector.}
\label{fig1}
\end{figure}

We first prepare the classical state $H$ and mix with the idler photon.
We use two single-photon detectors (NuCrypt CPDS-4), operating at 50$\,$MHz, to measure coincidences and accidental-coincidences between the idler and signal photons. A total of 16 settings are used for the HWPs and QWPs in the $\rm{PA_{s}}$ and $\rm{PA_{i}}$.
For each setting, we measure the coincidence counts (CC) and accidental-coincidence (AC) with an integration time of 2 seconds, corresponding to 100M sampling gates.
Using the maximum likelihood estimation, we reconstruct the $4 \times 4$ density matrix of the final state $\rho^{f}_{X}$ in the HV basis.
We repeat the state tomography for different classical states, $A$ and $L$, mixed with the idler photon.
We then define the numerically modeled final state $\rho^{n}_{X}$ as
\begin{equation}
\rho^{n}_{X}(C) = (1-x) \rho_{X}(C) + x\cdot 2(\mathcal{E}\rho_{w}(C)\mathcal{E}^{\dagger})
\label{eq:01}
\end{equation}
where $\rho_{X}(C)$ represents the quantum component of the measured $\rho^{f}_{X}$ state, and the $C = \{H, A, L\}$ denotes the injected classical state.
The state $\rho_{w}(C)$ is a Werner state, which is shared by the photon pairs before mixing the classical light in the idler channel. We obtain the Werner state $\rho_{w}(C)$ by subtracting the accidental-coincidences from the coincidences in the state tomography measurement. In our detection system, a coincidence count (CC) is recorded when both avalanche photodiodes (APDs) detect photons within the same gated time interval, whereas an accidental-coincidence count (AC) is recorded when both APDs detect photons in adjacent gated time intervals. The coincidence counts arise from both photon pairs and uncorrelated photons, while the accidental coincidence counts originate solely from uncorrelated photons. These uncorrelated photons mainly come from: (i) Raman photons generated by spontaneous Raman scattering in the optical fiber; (ii) loss of one photon from a pair in either the signal or idler channel; and (iii) injected classical light in the idler channel. By subtracting the accidental-coincidences from the coincidence counts (i.e., $\rm{CC - AC}$), we obtain the true coincidence counts arising from photon pairs. We use these true coincidence counts from the 16 measurement settings of the half-wave plate (HWP) and quarter-wave plate (QWP) in the quantum state tomography (QST) to reconstruct the Werner state.
We could measure the Werner state by blocking the weak coherent light. However, we avoid performing two separate QST measurements on this system. The reason is the accidental-subtracted Werner state has a probability identical to that of the Werner state measured when the weak coherent light is blocked.
The term $2(\mathcal{E}\rho_{w}(C)\mathcal{E}^{\dagger})$ is the classical$-$quantum state. The parameter $x$ denotes the small contribution from the classical$-$quantum state. The operator $\mathcal{E}=I\bigotimes\mathcal{B}$  indicates that $\mathcal{B}$ acts on the idler channel, while the identity operator  $I$  acts on the signal channel, since the signal photon is not mixed with the classical light. The symbol $\bigotimes$ denotes the tensor product. Consequently, the accidental-subtracted Werner state is transformed as $2(\mathcal{E}\rho_{w}(C)\mathcal{E}^{\dagger})$, resulting in a classical-quantum state, which is a $4\times4$ matrix.

We adopt the quantum discord framework to calculate the classical correlation of an open bipartite system, in which the environment performs a projective (von Neumann) measurement on the idler channel. The operator  $\mathcal{B}$ represents the outcome of the von Neumann projection on the idler channel and can be expressed as a linear combination of the Identity matrix and Pauli matrices, which are $2\times2$ matrices\cite{Luo_pra_2008,Luo_2008}. When $\mathcal{B}=H$, the polarization of the idler photon is projected onto the horizontal ($H$) state. Similarly, when $\mathcal{B}=A$ and $\mathcal{B}=L$, the polarization of the idler photon is projected onto the anti-diagonal ($A$) and left-circular ($L$) polarization states, respectively.
In our scenario, the weak coherent light increases the accidental-coincidences and induces the classical state, which can be interpreted as local measurement $\mathcal{B}$ on the idler photon. Therefore, the operator $\mathcal{B}$ is a $2\times 2$ density matrix representing the classical state $C$. The quantum discord for the classical-quantum state is zero.
We determine $\rho_{X}(C)$ and the parameter $x$ by solving the diagonal and off-diagonal density elements of the measured density matrix $\rho^{f}_{X}$ with those of $\rho^{n}_{X}(C)$, as defined in Eq.(1). \remove{There exists a unique solution in solving Eq.(1) that determines the operator $\mathcal{B}$.}\add{Eq.(1) admits a single admissible solution for the operator $\mathcal{B}$. Specifically, our method identifies a single polarization state from the discrete set of polarization states ($H$, $V$, $D$, $A$, $R$, $L$), and does not extend to superpositions of these states.} The resulting $\mathcal{B}$ is consistent with the classical state $C$ prepared by the weak coherent light.

The Bell-diagonal state\cite{Luo_pra_2008,BellDiagonal_2010,XMing_2011} can be written in matrix form as,
\begin{equation}
  \rho_{BD} =
  \begin{pmatrix}

    \frac{1+c_3}{4} & 0 & 0 & \frac{c_1-c_2}{4}\\
    0 & \frac{1-c_3}{4}& \frac{c_1+c_2}{4}& 0\\
    0 & \frac{c_1+c_2}{4} & \frac{1-c_3}{4} & 0\\
    \frac{c_1-c_2}{4} & 0 & 0 & \frac{1+c_3}{4}
 \end{pmatrix}
 \label{eq:02}
\end{equation}
where $c_{1,2,3}$ are the real parameters. The Werner state is obtained by setting $c_3 = p$, $c_1 = p$, and $c_2 = -p$. The Werner state is a mixture of the maximally entangled state $|\psi_{\circ}\rangle$ with the probability $p_w$ and the maximally mixed state as given by \cite{Werner_pra_1989,Munro_pra_2001},
\begin{eqnarray}
\rho_w = p_w|\psi_{\circ} \rangle \langle \psi_{\circ}| + \frac{1-p_w}{4} \mathcal{I}.
\label{eq:03}
\end{eqnarray}
where the $\mathcal{I}$ is the $4\times4$ identity matrix.
The Werner state can be represented as a $4 \times 4$  matrix:
\begin{equation}
  {\rho}_{w}=
  \begin{pmatrix}
     \frac{1+p_w}{4} & 0 & 0 & \frac{p_w}{2}\\
    0 & \frac{1-p_w}{4}& 0& 0\\
    0 & 0 & \frac{1-p_w}{4} & 0\\
    \frac{p_w}{2} & 0 & 0 & \frac{1+p_w}{4}
 \end{pmatrix}
 \label{eq:04}
\end{equation}
We calculate the probability $p_{w} =\sum^{n=6}_{i} p'_{i}/n $, where $p'_{1(2)} = 4 {\rho_w}_{11(44)}-1$, $p'_{3(4)} = 2{\rho_w}_{14(41)}$, and $p'_{5(6)}=1-4{\rho_w}_{22(33)}$. The density matrix elements ${\rho_w}_{11}$, ${\rho_w}_{44}$, ${\rho_w}_{14}$, ${\rho_w}_{41}$, ${\rho_w}_{22}$, and ${\rho_w}_{33}$ are obtained from the accidental-subtracted Werner state.
In this work, we consider the $X-$state $\rho_{X}$ of the form
\begin{equation}
  \rho_{X} =
  \begin{pmatrix}

    \frac{1+p}{4} & 0 & 0 & \frac{p}{2}\\
    0 & \frac{1-p}{4}&\rm{a} & 0\\
    0 & \rm{a}^{\ast} & \frac{1-p}{4} & 0\\
    \frac{p}{2} & 0 & 0 & \frac{1+p}{4}
 \end{pmatrix}.
 \label{eq:03}
\end{equation}
The state is referred to as an $X-$ state because the pattern of its density matrix elements ($p^{X}_{11}$, $p^{X}_{22}$, $p^{X}_{33}$, $p^{X}_{44}$, $p^{X}_{23}$, $p^{X}_{32}$, $p^{X}_{14}$, $p^{X}_{41}$) resembles the shape of the letter $X$. The basic properties of these real-valued parameters can be found in Ref\cite{Donnert_nmeth_2007}.
The $X-$state can be approximated as being close to a Werner state by setting the $c_3 = p$, $c_1 = 2 a + p$, and $c_2 = 2 a - p$, where $a$ is a small real number and much less than $p$, such that $c_1 \sim -c_2$.

In general, the local measurement operator $\mathcal{B}$ is extremely weak and therefore experimentally undetectable. To simulate its action, we inject classical light with a well-defined polarization state ($H$, $A$, or $L$) into the idler channel. In the following sections, we first demonstrate the validity of this approach by: (i) increasing the intensity of the classical light to approximately 1.5 times that of the idler photon, (ii) preserving a coincidence-to-accidental ratio (CAR) greater than 1 (or equivalently, $\rm{CC} - \rm{AC} > 0$), and (iii) preparing a simple horizontal ($H$) polarization state.
We then increase the intensity of the classical field for the $A$ and $L$ polarization states.
We want to demonstrate that, as the intensity of the classical field increases, our approach remains robust against (i) experimental imperfections (i.e., errors in  $\rm{CC} - \rm{AC}$ subtraction) and (ii) limitations of the experimental apparatus (i.e., the extinction ratio of the PBS in the $PA_i$ ). Even under these imperfect conditions, our approach is still capable of correctly identifying the corresponding classical polarization states.

Increasing the intensity of the classical light reduces the probability of the final state and its quantum component. However, as discussed above, the intensity does not affect the classical-quantum (CQ) state because the Werner state is obtained from the true coincidences. The polarization state of the classical light determines the operator $\mathcal{B}$.
The parameter $p_w$ represents the probability of the accidental-subtracted Werner state. The probability of the quantum component $\rho_X$ of the final state is reduced by the factor $c$, such that it becomes $p = c \times p_w$. The parameter $x$, with $1> x > 0$, denotes the contribution of the CQ state to the final state. The parameter $a$ is introduced in the $X-$state in Eq.(5), making it slightly different from the Werner state (which corresponds to $a=0$). The $X-$state in Eq.(5) is therefore a "Werner-like" $X-$state introduced in this work.
The density matrix elements of $\rho^{n}_{X}(C)$ in Eq.(1) are complex functions of the parameters $c$, $p_w$, $x$, and $a$. For simplicity, we use experimental data to demonstrate the algorithm used to solve Eq.(1).

\subsection{2.1 Procedure to identify Operator $\mathcal{B}$}
As mentioned above, the operator $\mathcal{B}$ used in $2(\mathcal{E}\rho_{w}(C)\mathcal{E}^{\dagger})$ in Eq.(1) corresponds to the polarization state of the injected classical light. For each choice of $\mathcal{B}$, we construct the corresponding $4\times4$ matrix of the classical-quantum (CQ) state. For each CQ state, specific density-matrix elements (real and imaginary parts) can be identified and used as constraints (i.e., greater or less than zero) in solving Eq.(1).

The procedure used to identify the operator $\mathcal{B}$ is outlined as follows:

\indent 1. We perform quantum state tomography (QST) on the final state and reconstruct its density matrix.

\indent 2. Using the true coincidence counts from the QST data, we reconstruct the $4\times4$ matrix of the accidental-subtracted Werner state.

\indent 3. The probability $p_w$ is then obtained from the accidental-subtracted Werner state.

\indent 4. We solve Eq.(1) by (i) using the density-matrix elements of the measured final state and (ii) imposing the constraints derived from each candidate $\mathcal{B}$ operator in the corresponding CQ state. Since we have no prior knowledge of the polarization state of the classical field (i.e., the operator $\mathcal{B}$), we perform Mathematica function NSolve separately for each possible $\mathcal{B}$, using the constraints obtained by its corresponding CQ state.

\subsection{2.2 Detection of Classical State H}

We first measure the coincidence and accidental-coincidence counts of the correlated photon pairs without the weak coherent light. These correlated photon pairs are generated via the CPS scheme using a horizontally polarized pump. We measure coincidence-to-accidental ratio (CAR) of the signal and idler photons after $PA_{s}$ and $PA_{i}$, both set to horizontal (H) polarization. We obtain the CAR of 4. Then, we inject weak coherent light with the H polarization state, along with the H-polarized idler photons, into $PA_{i}$. The average power of the weak coherent light is adjusted so that the single counts of the idler photons increase by a factor of 1.5.
We observe that the CAR drops from 4 to 3. In a later section, we increase the single counts of the idler photon by factors of 10 and 20 for the detection of classical states $A$ and $L$, respectively.

We generate a two-photon polarization-entangled state from the CPS and perform the QST with the weak coherent light in the idler channel. We reconstruct the density matrix of the final state $\rho^{f}_{X}(H)$ (see the Methods for the measured density matrix of the classically mixed idler and signal photons as shown in Eq.(17)).
For solving Eq.(1), we use the measured density matrix elements of the final state, $\rho^{f}_{X}(H)$ : $p^{f}_{11}$, $p^{f}_{14}$, $p^{f}_{23}$ and $p^{f}_{13}$.
The measured final state has the same structure as the $X-$state given in Eq.(5), which can ideally be described by two parameters, $p$ and $a$. However, the experimentally measured final state is not perfect; for example, the density matrix elements $p^{f}_{11}$ and $p^{f}_{44}$ are not equal. We therefore select  $p^{f}_{11}$, $p^{f}_{14}$, $p^{f}_{23}$ and $p^{f}_{13}$ from the measured final state to solve the Eq.(1). This approach allows us to obtain a numerically solved final state whose density matrix elements closely match those of the measured final state, i.e., $p^{n}_{44} \sim p^{f}_{44}$, etc.
If $p^{f}_{44}$ is included in addition to the four selected density matrix elements, Eq.(1) cannot be solved because the experimentally measured values $p^{f}_{11}$ and $p^{f}_{44}$ are not equal. Therefore, choosing these four density matrix elements provides the optimal and consistent set of constraints for solving Eq.(1).

We obtain the Werner state $\rho_{w}(H)$  by subtracting the accidental-coincidences from the coincidences. The Werner state $\rho_{w}(H)$  is given by,
\begin{equation}
  \rho_{w}(H) =
  \begin{pmatrix}
    0.499 & 0.007 & 0.00258 & 0.471\\
    0.007 & 0.000466&-0.00087 & 0.00544\\
    0.00258 & -0.00087& 0.00525 & 0.017\\
    0.471 & 0.00544 & 0.017 & 0.495
 \end{pmatrix}.
 \label{eq:04}
\end{equation}
The imaginary part of $\rho_{w}(H)$ is close to 0. We obtain the probability of the Werner state, $p_w$, as 0.972.
We then compute, $2(\mathcal{E}\rho_{w}(H)\mathcal{E}^{\dagger})$, which is given by;
\begin{equation}
  2(\mathcal{E}\rho_{w}(H)\mathcal{E}^{\dagger}) =
  \begin{pmatrix}
    0.998 & 0.00 & 0.00516 & 0.00\\
    0.00 & 0.00&0.00 & 0.00\\
    0.00516 & 0.00& 0.0105 & 0.00\\
    0.00 & 0.00 & 0.00 & 0.00
 \end{pmatrix},
 \label{eq:05}
\end{equation}
where we choose $\mathcal{B}=H = \begin{pmatrix}
    1 & 0 \\
    0 & 0
  \end{pmatrix}$  and the identity matrix $I=\begin{pmatrix}
    1 & 0 \\
    0 & 1
  \end{pmatrix}$.
The off-diagonal elements of $2(\mathcal{E}\rho_{w}(H)\mathcal{E}^{\dagger})$, i.e., $p_{13}$ and $p_{31}$ are positive, and these conditions are used to solve Eq.(1).
We substitute $2(\mathcal{E}\rho_{w}(H)\mathcal{E}^{\dagger})$ into the Eq.(1). (See the Methods for the other scenario $2(\mathcal{E'}\rho_{w}(H)\mathcal{E'}^{\dagger})$, where $\mathcal{B}$ operates on the signal photon and $\mathcal{E'}=\mathcal{B}\bigotimes\mathcal{I}$. The same procedure is applied for the classical state $V$, where $\mathcal{B}=V$, in both the $\mathcal{E}$ and $\mathcal{E'}$ cases.)
We replace $p $ with  $c \cdot p_w$ in  $\rho_{X}(H)$ from Eq.(1). This gives the density matrix $\rho^{n}_{X}(H)$ from Eq.(1) in terms of the parameters $c$, $a$, and $x$.
We then solve the elements of $\rho^{n}_{X}(H)$ ,i.e., $p^{n}_{11}$, $p^{n}_{14}$, and $p^{n}_{23}$, with the corresponding elements of the measured final $X$-state density matrix: $p^{f}_{11}$, $p^{f}_{14}$, and $p^{f}_{23}$. We also impose the condition $p^{n}_{13} > 0$ (see the Methods for details on this condition. One can impose additional conditions, which are chosen based on the related classical-quantum states). Specifically, we set  $p^{n}_{11}=p^{f}_{11}=0.416$, $p^{n}_{14}=p^{f}_{14}=0.282$, $p^{n}_{23}=p^{f}_{23}=-0.0078$, with $p^{n}_{13}=p^{f}_{13} > 0$. We numerically solve this system by using the Mathematica function $\texttt{NSolve}$ as shown in the Methods.

We obtain the density matrix $\rho^{n}_{X}(H)$ of Eq.(1), given by:
\begin{equation}
  \rho^{n}_{X}(H) =
  \begin{pmatrix}
    0.416 & 0.00 & 0.0001 & 0.282\\
    0.00 & 0.100&-0.00785 & 0.00\\
    0.0001 & -0.00785& 0.1009 & 0.00\\
    0.282 & 0.00 & 0.00 & 0.382
 \end{pmatrix}.
 \label{eq:06}
\end{equation}
The $X-$state density matrix elements of $\rho^{n}_{X}(H)$ are comparable to those of the measured final state $\rho^{f}_{X}(H)$ as shown in Eq.(17) in the Methods section.
We obtain the parameters $x$ =0.0334, $c$=0.5997, and $a$=-0.0081. The probability of $\rho_{X}(H)$ in Eq.(1) is given by $c \times p_w = 0.5997 \times 0.972 = 0.582$. This relatively low probability results from mixing the weak coherent light with the idler photon, which increases accidental-coincidences in the state tomography measurement. The parameter $x$ is only $3.0\%$, indicating a small contribution from the classical$-$quantum state to the final state $\rho^{n}_{X}(H)$. However, this contribution is sufficient to detect the classical state $H$ in the idler channel. If we choose other operators $\mathcal{B}= (V, D, A, R, L)$ to compute $2(\mathcal{E}\rho_{w}(H)\mathcal{E}^{\dagger})$ in Eq.(1), the Eq.(1) cannot be solved, and the parameter $x$ cannot be obtained.
For example, if we assume $\mathcal{B}=A$ and impose the condition $p^{n}_{\rm{off\,diag.}}$ =$p^{f}_{\rm{off\,diag.}}<0 $
(i.e., $p^f_{12}$, $p^f_{21}$, $p^f_{13}$, $p^f_{31}$, $p^f_{24}$, $p^f_{42}$, $p^f_{34}$, $p^f_{43}$$) < 0$
(as discussed in Section 2.3), Eq.(1) has no solution. This is because the experimentally measured off-diagonal elements $p^{f}_{\rm{off\,diag.}}$ of the final state (see Eq.(17) in Section 5.1) are not all negative. Similarly, for $\mathcal{B}=V$, $L$, and other polarization states, Eq.(1) has no solution, whereas a consistent solution exists only for $\mathcal{B}=H$.
This procedure therefore allows us to identify the polarization state of the injected classical light.
Since the parameter $x$ is small, the $\rho^{n}_{X}(H)$ is primarily contributed from its quantum part, $\rho_{X}(H)$.

\subsection{2.3 Detection of Classical State A}

Using a similar approach to that in the previous section, we inject the classical state $A$ into the idler channel.  We observe that the single counts of the idler channel increase by a factor of 10, and the CAR decreases from 4 to 1.
We reconstruct the density matrix of the measured final state (see the Methods for the measured density matrix $\rho^{f}_{X}(A)$ in Eq.(18)).
The accidental-subtracted Werner state $\rho_{w}(A)$ is obtained as;
\begin{equation}
  \rho_{w}(A) =
  \begin{pmatrix}
    0.488 & 0.0157 & -0.00995 & 0.431\\
    0.0157 & 0.00208&-0.00117 & 0.0212\\
    -0.00995 & -0.00117& 0.0103 & 0.000182\\
    0.431 & 0.0212 & 0.000182 & 0.499
 \end{pmatrix},
 \label{eq:07}
\end{equation}
where the imaginary part of $\rho_{w}(A)$ can be neglected. The probability of the Werner state, $p_w$, is 0.937. This is lower than in the previous case because more classical photons are injected in the idler channel and imperfect $\rm{CC - AC}$ subtraction.
The state $2(\mathcal{E}\rho_{w}(A)\mathcal{E}^{\dagger})$ is then computed as:
\begin{equation}
  2(\mathcal{E}\rho_{w}(A)\mathcal{E}^{\dagger}) =
  \begin{pmatrix}
    0.2293 & -0.2293 & -0.2092 & 0.2092\\
    -0.2293 & 0.2293&0.2092 & -0.2092\\
   -0.2092 & 0.2092& 0.2544 & -0.2544\\
    0.2092 & -0.2092 & -0.2544 & 0.2544
 \end{pmatrix},
 \label{eq:08}
\end{equation}
where we choose $\mathcal{B}=A = \begin{pmatrix}
    \frac{1}{2} & \frac{-1}{2} \\
    \frac{-1}{2} & \frac{1}{2}
  \end{pmatrix}$.
The formation of non-zero density matrix elements in this  $2(\mathcal{E}\rho_{w}(A)\mathcal{E}^{\dagger})$ is significantly different from that $2(\mathcal{E}\rho_{w}(H)\mathcal{E}^{\dagger})$ in Eq.(7). This is because the classical state $A$ has the $H$ and $V$ components.  The off-diagonal elements of $2(\mathcal{E}\rho_{w}(A)\mathcal{E}^{\dagger})$, i.e., $p^{X}_{\rm{off\,diag.}}$= $\{$$p_{12}$, $p_{21}$, $p_{13}$, $p_{31}$, $p_{24}$, $p_{42}$, $p_{34}$, $p_{43}$$\}$, are all negative, and these conditions are used to solve Eq.(1).
We substitute $2(\mathcal{E}\rho_{w}(A)\mathcal{E}^{\dagger})$ into Eq.(1). Additional details, including the alternative transformation $2(\mathcal{E'}\rho_{w}(A)\mathcal{E'}^{\dagger})$, i.e., $\mathcal{B}$ operates on the signal photon with $\mathcal{E'}=\mathcal{B}\bigotimes\mathcal{I}$, as well as the case for the classical state $D$, where $\mathcal{B}=D$, are provided in the Methods.

We replace $p $ with $c \times p_w$ in  $\rho_{X}(A)$ from Eq.(1), and obtain the density matrix $\rho^{n}_{X}(A)$ in terms of parameters $c$, $a$, and $x$. We solve the elements of $\rho^{n}_{X}$, specifically, $p^{n}_{11}$, $p^{n}_{14}$, and $p^{n}_{23}$, with those of the measured $X$-state density matrix: $p^{f}_{11}$, $p^{f}_{14}$, and $p^{f}_{23}$. We also impose the condition $p^{n}_{\rm{off\,diag.}} < 0$ (See the Methods section for details on imposing this condition, which are chosen based on the corresponding classical-quantum states). We set  $p^{n}_{11}=p^{f}_{11}=0.349$, $p^{n}_{14}=p^{f}_{14}=0.206$, $p^{n}_{23}=p^{f}_{23}=0.0219$, with $p^{n}_{\rm{off\,diag.}} = p^{f}_{\rm{off\,diag.}} < 0$. We numerically solve Eq.(1) by using the Mathematica function $\texttt{NSolve}$ as shown in the Methods.

We obtain the density matrix $\rho^{n}_{X}(A)$ from Eq.(1) as:
\begin{equation}
  \rho^{n}_{X}(A) =
  \begin{pmatrix}
    0.349 & -0.007 & -0.0067 & 0.206\\
    -0.007 & 0.149&0.0219 & -0.0066\\
    -0.0067 & 0.0219& 0.1504 & -0.0081\\
    0.206 & -0.0066 & -0.0081 & 0.349
 \end{pmatrix}
 \label{eq:09}
\end{equation}
The $X-$state density matrix elements of $\rho^{n}_{X}(A)$ are comparable to those of the measured final state $\rho^{f}_{X}(A)$ as shown in Eq.(16) in the Methods section.
We obtain the parameters $x =0.03192$, $c=0.43943$, and $a=0.01572$. The probability of $\rho_{X}(A)$ in Eq.(1) is given by $c \times p_w = 0.43943 \times 0.937 = 0.411$.
The parameter $x$ is still only $3.0\%$, indicating a small contribution of the classical-quantum state to the final state $\rho^{f}_{X}(A)$.
If we choose other operators $\mathcal{B}=(H, V, D, R, L)$ for $2\mathcal{E}\rho_{w}(A)\mathcal{E}^{\dagger}$ in Eq.(1), we are unable to solve the system.
The classical state $A$ is the only viable solution for $\rho^{n}_{X}(A)$ in Eq.(1). The small value of $x$ indicates that $\rho^{n}_{X}(A)$ is mainly contributed by its quantum component, $\rho_{X}(A)$.

\subsection{2.4 Detection of Classical State L}

We inject the classical state $L$ into the idler channel and increase the power of the weak coherent light until the single counts in the idler channel rise by a factor of 20. As a result, the CAR decreases from 4 to 1.
We perform the quantum state tomography and reconstruct the density matrix of the measured final state (see the Methods section for the measured density matrix $\rho^{f}_{X}(L)$ in Eq.(19)).

Similarly, we use the density matrix elements of the measured final state $\rho^{f}_{X}(A)$: $p^{f}_{11}$, $p^{f}_{14}$, $p^{f}_{23}$ and the condition $\rm{Im}[p^{f}_{12}] < 0$ and $\rm{Im}[p^{f}_{13}] > 0$, to solve Eq.(1).
We obtain the accidental-subtracted Werner state $\rho_{w}(L)$ as follows:
\begin{equation}
  \rho_{w}(L) =
  \begin{pmatrix}
    0.505 & -0.0298 & 0.0217 & 0.405\\
    0.0298 & 0.00966&-0.00266 & -0.0077\\
    0.0217 & -0.00266& 0.0101 & 0.0464\\
    0.405 & -0.00767 & 0.0464 & 0.474
 \end{pmatrix}.
 \label{eq:10}
\end{equation}
We find the probability of the Werner state to be $p_w = 0.909$, which is expected due to the increased number of classical photons injected in the idler channel, imperfection in subtracting $\rm{CC - AC}$,  and the extinction ratio of the PBS in the $PA_{i}$.
We then compute $2(\mathcal{E}\rho_{w}(L)\mathcal{E}^{\dagger})$ as follow:
\begin{equation}
  \begin{pmatrix}
  0.25733 & 0.0-i\,0.2573 & 0.007+i\,0.2038 & 0.2038-i\,0.007\\
  0.0+i\,0.2573 & 0.25733&-0.2038+i\,0.007 & 0.007+i\,0.2038\\
  0.007-i\,0.2038 & -0.2038-i\,0.007& 0.2421 & 0.0-i\,0.2421\\
  0.2038+i\,0.007 & 0.007-i\,0.2038 & 0.0+i\,0.2421 & 0.2421
 \end{pmatrix},
 \label{eq:11}
 \end{equation}
where we choose $\mathcal{B}=L = \begin{pmatrix}
    \frac{1}{2} & \frac{i}{2} \\
    \frac{-i}{2} & \frac{1}{2}
  \end{pmatrix}$.
We observe the formation of non-zero imaginary numbers in $2(\mathcal{E}\rho_{w}(L)\mathcal{E}^{\dagger})$  because the classical state $L$ has an imaginary component $V$. The imaginary parts of off-diagonal elements of $2(\mathcal{E}\rho_{w}(L)\mathcal{E}^{\dagger})$ are used as conditions to solve Eq.(1).
We substitute $2(\mathcal{E}\rho_{w}(L)\mathcal{E}^{\dagger})$ into Eq.(1). (See the Methods for $\mathcal{B}$ operates on the signal photon, i.e., $2(\mathcal{E'}\rho_{w}\mathcal{E'}^{\dagger})$, where $\mathcal{E'}=\mathcal{B}\bigotimes\mathcal{I}$. Also refer to the case of the classical state $R$, where $\mathcal{B}=R$, for both $\mathcal{E}$ and $\mathcal{E'}$.)
We replace $p$ with $c \times p_w$ in  $\rho_{X}(L)$ from Eq.(1). We then obtain the density matrix $\rho^{n}_{X}(L)$ in terms of $c$, $a$, and $x$.
We set  $p^{n}_{11}=p^{f}_{11}=0.275$, $p^{n}_{14}=p^{f}_{14}=0.0535$, $p^{n}_{23}=p^{f}_{23}=-0.0011$, with $\rm{Im}[p^{n}_{12}]=\rm{Im}[p^{f}_{12}] < 0$ and $\rm{Im}[p^{n}_{13}]=\rm{Im}[p^{f}_{13}] > 0$  (See the Methods section for details on these conditions. Additional conditions can be imposed as necessary, chosen based on the related classical$-$quantum states).
We numerically solve Eq.(1) by using the Mathematica function $\texttt{NSolve}$ as shown in the Methods.

We numerically solve Eq.(1) and obtain the density matrix $\rho^{n}_{X}(L)$ as:
\begin{equation}
  \begin{pmatrix}
    0.275 & 0.0015-i\,0.0037 & 0.0+i\,0.004 & 0.053+i\,0.0001\\
    0.0015+i\,0.0037 & 0.2249&-0.0011+i\,0.0001 & 0.0+i\,0.004\\
    0.0-i\,0.004 & -0.0011-i\,0.0001& 0.2253 & 0.0015-i\,0.0037\\
    0.053-i\,0.0001 & 0.0-i\,0.004 & 0.0015+i\,0.0037 & 0.274
 \end{pmatrix},
 \label{eq:06}
\end{equation}
where the $X-$ state density matrix elements closely match those of the measured final state $\rho^{f}_{X}(L)$ as shown in Eq.(19) in the Methods section.
We obtain the parameters $x =0.0185$, $c=0.111$, and $a=0.0027$. The probability of the state $\rho_{X}(L)$ in Eq.(1) is given by $c \times p_w = 0.111 \times 0.909 = 0.10$.
The parameter $x$ is only $1.8\%$, which is smaller than that of the classical states $H$ and $A$, yet still we successfully detect the classical state $L$ in the idler channel.

In summary, when we increase the intensity of the classical light by factors of 1.5, 10, and 20 for the $H$, $A$, and $L$ polarization states, respectively, the corresponding values of the parameter $x$ are 0.03, 0.03, and 0.018. These results indicate only modest variations in $x$, with an average value of approximately 0.026 (about 3$\%$).
Increasing the intensity of the classical light leads to a corresponding increase in accidental counts for the $H$, $A$, and $L$ cases. As a result, the probability $c\times p_w$ of the quantum component of the final state decreases accordingly. This demonstrates that our method remains effective in a robust environment where the quantum discord can be very small, as discussed in the following section.

\subsection{2.5 The relative strength between $p^{f}_{11}$ and $p^{f}_{14}$}

It is important to explore the relative strength of $p^{f}_{11}$ and $p^{f}_{14}$ in solving Eq.(1), under specific conditions on the parameters $a$ and $p^{f}_{23}$. For example, let us define $y=\frac{p^{f}_{11}}{p^{f}_{14}}$ and substitute $\mathcal{B}=A$ in Eq.(1). The matrix elements of $\rho^{n}_{X}(A)$ are functions of $x$, $p$, and $a$. We impose conditions  $p^{n}_{12}=p^{f}_{12}<0$ and $p^{n}_{23}=p^{f}_{23}>0$, to solve the equality $y=\frac{p^{f}_{11}}{p^{f}_{14}}$, subject to the constraints $1>x>0$, $1>p>0$, $1>y>0$, and $a <0$ or $a > 0$. We obtain the expression $y =\frac{p\cdot x -2\cdot p}{p\cdot x -p-1}$ for positive values of $a > 0 $, and negative values of $a$ in the range $\frac{p\cdot x}{4\cdot x -4}<a<0$ as shown in Fig.2(a) and in the top region of Fig.2(b).
This indicates that the parameter $y$ exists with certain negative values of $a$. If we change the detection of the classical state to $D$, we substitute $\mathcal{B}=D$ in Eq.(1). We impose the conditions $p^{n}_{12}=p^{f}_{12}>0$ and $p^{n}_{23}=p^{f}_{23}<0$ when solving the equality $y=\frac{p^{f}_{11}}{p^{f}_{14}}$.  We obtain the parameter $y =\frac{p\cdot x -2\cdot p}{p\cdot x -p-1}$ for negative values of $a < \frac{p\cdot x}{4\cdot x -4.0}$ as described in Fig.2(a) and in the bottom region of Fig.2(b).
There is no solution for the parameter $y$ associated with positive values of $a > 0$. If the reconstructed final states for the classical states $A$ and $D$ do not satisfy the parameters $y$ and $a$, as shown in Fig. 2 (a) and 2(b), we will be unable to solve Eq.(1).

\begin{figure}
 \centering
        \includegraphics[width=1.0\textwidth]{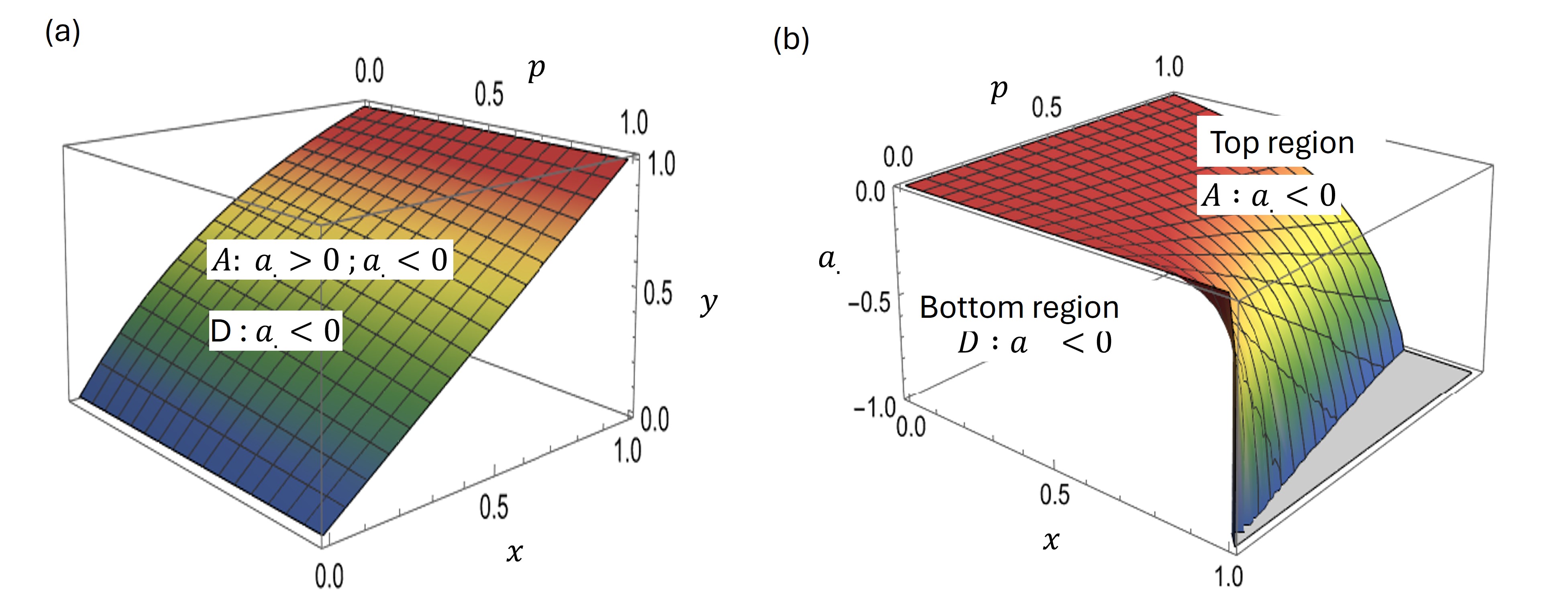}
 \caption{(a). The plot of $y =\frac{p\cdot x -2\cdot p}{p\cdot x -p-1}$ for the classical state $A$ whether $a <0$ or $a > 0$. The same plot applies to the classical state $D$ for the negative values of $a < \frac{p\cdot x}{4\cdot x -4.0}$. (b). The plot of $a =\frac{p\cdot x}{4\cdot x -4.0}$. The top region is in the range $\frac{p\cdot x}{4\cdot x -4}<a<0$ for the classical state $A$. The bottom region is in the range $a < \frac{p\cdot x}{4\cdot x -4}$ for the classical state $D$.}
\label{fig1}
\end{figure}

\subsection{2.6 Quantum Discord}

It is important to verify that the state $\rho_{w}$, shared by two parties (Alice and Bob), has non-zero quantum discord before calculating the classical-quantum state $2(\mathcal{E}\rho_{w}\mathcal{E}^{\dagger})$. We believe that a high value of quantum discord is necessary for our classical state detection method. We directly calculate the diagonal quantum discord of $\rho_{w}$ and the quantum component of the final $X-$ state $\rho^{n}_{X} $. 
The diagonal quantum discord is given by\cite{SLloyd_2019,SLloyd_2015}:
\begin{equation}
\mathcal{D}_{Q}(\rho_{w}) = \mathcal{I}(\rho_{w})-\mathcal{I}(\pi(\rho_{w})),
\label{eq:13}
\end{equation}
where $\mathcal{I}$ is total correlation,
$\pi(\rho_{w})=\mathcal{E}_{\mathcal{B}}\rho_{w}\mathcal{E}_{\mathcal{B}}^{\dagger}+\mathcal{E}_{\mathcal{B'}}\rho_{w}\mathcal{E}_{\mathcal{B'}}^{\dagger}$, and
$\mathcal{E}_{\mathcal{B}}+\mathcal{E}_{\mathcal{B'}}=I$, where $\mathcal{B}=(H, A, L)$ and $\mathcal{B'}=(V, D, R)$, respectively.
We obtain the diagonal quantum discord values $\mathcal{D}_{Q}(\rho_{w}(H))=\rm{0.8373}$, $\mathcal{D}_{Q}(\rho_{w}(A))=\rm{0.9338}$, and $\mathcal{D}_{Q}(\rho_{w}(L))=\rm{0.9210}$. This indicates that the diagonal quantum discords for the projection operators $A$ and $L$ are higher than that for the operator $H$.
We obtain similar quantum discord values using the formulation prior to the optimization of classical correlation, as outlined in Ref.~\cite{Luo_pra_2008}. In addition, we verify that the quantum discords for the classical-quantum states in Eq.(7), Eq.(10), and Eq.(13) are zero.

We also calculate the general quantum discord $\mathcal{D}_{w}$ of a Werner state, where full optimization is used to estimate the classical correlations, as shown in Ref\cite{Luo_pra_2008}. The optimized $\mathcal{D}_{w}$ is a function of $p_w$, given by:
\begin{equation}
\mathcal{D}_{w}(\rho_{w})=\frac{1-p_w}{4}\log_{2}(1-p_w)-\frac{1+p_w}{2}\log_{2}(1+p_w)+\frac{1+3\cdot p_w}{4}\log_{2}(1+3\cdot p_w).
\label{eq:014}
\end{equation}
We obtain the quantum discord values  $\mathcal{D}_{w}(\rho_{w}(H))=\rm{0.927}$, $\mathcal{D}_{w}(\rho_{w}(A))=\rm{0.852}$, and $\mathcal{D}_{w}(\rho_{w}(L))=\rm{0.800}$. The quantum discord $\mathcal{D}_{w}$ decreases accordingly with decreasing $p_w$.
Both the diagonal quantum discord $\mathcal{D}_{Q}(\rho_{w})$ and general quantum discord $\mathcal{D}_{w}(\rho_{w})$ of the $\rho_{w}$ in Eq.(1) are above 0.8. This indicates that the quantum correlations of the accidental-subtracted state $\rho_{w}$ are sufficient for detecting classical states.

Finally, we calculate the quantum discord of the quantum components $\rho_{X}(H)$, $\rho_{X}(A)$, and $\rho_{X}(L)$ using the general quantum discord defined in Eq.(16). The quantum discord values for $\rho_{X}(H)$, $\rho_{X}(A)$, and $\rho_{X}(L)$ are 0.310, 0.184, and 0.013, respectively.
These quantum discord values correspond to their respective final $X-$ states, as the parameter x $\approx 0.03$ is small. The near-zero quantum discord of the $\rho_{X}(L)$ is attributed to the higher intensity of the weak coherent light in classical state $L$, compared to the classical states $A$ and $H$.

\section{3. Discussion}

We are able to detect the classical state injected into one photon from a photon-pair by performing full quantum state tomography. Our detection algorithm exploits the concept of LOCC and quantum discord, both of which are related to measurement-induced (environment-induced) correlations between the quantum system and the injected classical state. The strength of this correlation becomes small as the classical light increases the single counts of the idler photon by a factor up to 20. We obtain the parameter $x$ about 3.0$\%$, which represents the contribution of a classical-quantum state to the measured final state. A classical-quantum (quantum-classical) bipartite  state has zero quantum discord when LOCC is performed on the idler (or signal) photon, respectively. The positive and negative values of the real and imaginary parts of the off-diagonal elements of the classical-quantum state's density matrix are used to impose the conditions on $\rho^{n}_{X}$ in solving Eq.(1).

We perform local measurements $\mathcal{B}$ on the idler photon by injecting weak coherent light, and then investigate the pre- and post-measurement states. The pre-measurement state is the accidental-subtracted Werner state, and the post-measurement state corresponds the measured final state. The weak coherent light, with a well-defined polarization state, can be interpreted as a measurement apparatus because its intensity is higher  than that of the idler photon. As a result, the idler channel effectively becomes a classical channel with a well-defined polarization state after the measurement.

We chose the final state to be an $X-$ state, rather than a Werner state, because it represents a more general form of quantum state that undergoes decoherence. The injection of a classical state  can lead to separable states in a bipartite system shared by the idler and signal photons. These states can still exhibit non-classical correlations and possess non-zero quantum discord. We observe non-zero quantum discord for the quantum components $\rho_{X}(H)$, $\rho_{X}(A)$, and $\rho_{X}(L)$ of the final states. However, the quantum discord almost vanishes (i.e., drops to approximately 1.3$\%$) as the intensity of the weak coherent light increases. We also obtain diagonal quantum discord values greater than 80$\%$ for the accidental-subtracted Werner states, which are used in calculating the classical-quantum states. Ideally, one could obtain the initial Werner state by blocking the weak coherent light and then use it to compute the classical-quantum state. However, doing so would require performing two full quantum state  tomographies$\,-\,$one with and one without the weak coherent light$\,-\,$which makes the method impractical for real-time, long-distance quantum communication.

We can further reduce the coincidence measurement time. The detector efficiency in our experiment is approximately 20$\%$. We use 100M sampling gates, or 2 seconds, for each of the 16 settings of the HWPs and QWPs. The total integration time for QST is approximately 16 $\times$ 2 =32 seconds. If we use a superconducting single photon detector with an efficiency of up to 70$\%$$\,-\,$a factor of 3.5 improvement$\,-\,$then the coincident count would increase by a factor of 3.5$\times$3.5 =12.25. As a result, the total integration time could be reduced to $\frac{32}{12.25}=2.6 \,\rm{seconds}$. In addition, we demonstrate the method using a photon-pair source with a CAR as low as 4, and a Werner state probability $p_w$ as low as 0.90.

The existence of a viable solution to Eq.(1) in our algorithm relies on the properties of the $X-$ state and the off-diagonal elements of the classical-quantum state. However, the relative strength between $p^{f}_{11}$ and $p^{f}_{14}$, along with the sign of the parameter $a $ (i.e., whether $a > 0\, \rm{or}\, a < 0$) in the quantum component of the measured final state, are critical constraints for obtaining such a solution.  A solution to Eq.(1) may not be attainable if the measured final state does not satisfy these constraints, as discussed in previous sections.

In the classical-quantum coexistence scheme\cite{Gamze_2025}, classical light co-propagates with quantum light in an optical fiber, but at different optical wavelengths. The classical light can be used to monitor and compensate for the polarization drift experienced by the quantum light\cite{Gamze_Q2025}. The classical and quantum signals are separated by inserting a wavelength-division multiplexer (WDM) immediately before the polarization analyzer and single-photon detectors. To achieve effective polarization compensation for the quantum light, the wavelengths of the classical and quantum signals are ideally matched.
In a recent polarization-entangled photon distribution experiment over New York City fiber\cite{Mehdi_JOC_2025,Mehdi_PRX_2025}, classical light at the same wavelength as the quantum light was used to perform automated polarization compensation. However, during this compensation process, the quantum light was turned off.
In contrast, our method enables measurement of the polarization state of the classical light using QST, with both the classical and quantum signals passing through the polarization analyzer and into single-photon detector. We believe that this approach can improve polarization compensation in classical-quantum coexistence networks, such as quantum wrapping, where classical and quantum signals coexist in different time domains.

Finally, our algorithm can be incorporated into machines learning \cite{Guo_2025,Torlai_NP_2018} and quantum process tomography\cite{Chuang_JMO_1997,Altepeter_PRL_2003} with unsupervised learning\cite{Torlai_NC_2023} to detect quantum processes induced by the environment. Quantum processes or operators can be extracted from the process tomography, including complete projection measurements (i.e., von Neumann measurements) expressed in terms of the Pauli matrices ($\sigma_x$, $\sigma_y$, $\sigma_z$) and the identity matrix.

\section{4. Conclusion}

We prepare a weak coherent light with a well-defined polarization state and mix it with the idler photon from a polarization entangled photon pair. By performing full quantum state tomography, we demonstrate the detection of the classical state using the concepts of LOCC and classical-quantum states. The classical state detection model shows good agreement with the experimental results. Importantly, our approach does not require high fidelity entangled states, making it suitable for practical applications and robust against environmental decoherence in real-time, long-distance quantum communication. This method has the potential to significantly advance current classical-quantum coexistence networking techniques, such as quantum wrapping. Moreover,  our algorithm can support future developments in quantum key distribution (QKD), particularly for protocols that involve the coexistence of weak coherent states and entangled photon states.

\section{5. Methods}

\subsection{5.1 The Reconstructed Density Matrix of the measured final $X-$ States}.

The classical states $H,A,\rm{and}\, L$ are independently prepared from the weak coherent light. The classical state is then injected to the idler channel at the same optical wavelength of the idler photon via a DWDM.

The final state $\rho^{f}_{X}(H)$ is given by,
\begin{equation}
\scalebox{0.75}{$
  \begin{pmatrix}
    0.416 & \num{-4.3e-04} + i6.5\times 10^{-03} &  2.2\times10^{-03} - i2.7\times10^{-03} & 0.2 + i0.033\\
    \num{-4.3e-04} - i\num{6.5e-03} & 0.0858&\num{-7.8e-03} + i0.0364& \num{2.8e-03} - i\num{5.6e-03}\\
     \num{2.2e-03} + i\num{2.7e-03} & \num{-7.8e-03} - i0.0364&  0.122 & \num{-2.4e-04} + i\num{1.2e-03}\\
    0.282 - i0.033 &  \num{2.8e-03} + i\num{5.6e-03} & \num{-2.4e-04} - i\num{1.2e-03} & 0.374
 \end{pmatrix}.
$}
 \label{eq:M1}
\end{equation}

The final state $\rho^{f}_{X}(A)$ is given by,
\begin{equation}
\scalebox{0.75}{$
  \begin{pmatrix}
    0.349 & \num{-7.0e-04} + i0.0263 &  -0.113 + i0.0314 &  0.206 - i0.055\\
    \num{-7.0e-04} - i0.0263 & 0.132& 0.0219 - i0.0171& -0.0921 + i0.047\\
     -0.113 - i0.0314 & 0.0219 + i0.0171& 0.139 & \num{-2.47e-03} - i0.027\\
    0.206 + i0.055 &  -0.0921 - i0.047 & \num{-2.4e-03} + i0.027 & 0.378
 \end{pmatrix}.
 $}
 \label{eq:M2}
\end{equation}

The final state $\rho^{f}_{X}(L)$ is given by,
\begin{equation}
\scalebox{0.75}{$
  \begin{pmatrix}
    0.275 & -0.0114 - i\num{6.3e-03} &  -0.0669 + i0.199 &  0.0535 - i0.0172\\
    -0.0114 + i\num{6.3e-03} & 0.232& -\num{1.10e-03} - i\num{6.0e-03}& -0.0648 + i0.213\\
     -0.0669 - i0.199 & -\num{1.1e-03} + i\num{6.0e-03}&  0.223 &  \num{6.5e-03} - i\num{2.1e-03}\\
     0.0535 + i0.0172 &  -0.0648 - i0.213 & \num{6.5e-03} + i\num{2.1e-03} &  0.268
 \end{pmatrix}.
 $}
 \label{eq:M3}
\end{equation}

\subsection{5.2 The $2(\mathcal{E'}\rho_{w}(H)\mathcal{E'}^{\dagger})$ with the $\mathcal{E'}=(\mathcal{B=H})\bigotimes\mathcal{I}$}

\begin{equation}
\small
  2(\mathcal{E'}\rho_{w}(H)\mathcal{E'}^{\dagger}) =
  \begin{pmatrix}
    0.998 & 0.014&  0.0 &  0.0\\
    0.014  & 0.000932& 0.0& 0.0\\
    0.0 & 0.0&  0.0 &  0.0\\
     0.0 &  0.0 & 0.0 &  0.0
 \end{pmatrix}
 \label{eq:M3}
\end{equation}.
We choose $p_{12}=p_{21} > 0$ and impose $p^{n}_{12} > 0$ as a condition for solving Eq.(1).

\subsection{5.3 The $\mathcal{E}$ and $\mathcal{E'}$ for $\mathcal{B}=V$}
Here we can assume $\rho_{w}(V)=\rho_{w}(H)$ because the Werner state is accidental subtracted.
\begin{equation}
\small
  2(\mathcal{E}\rho_{w}(V)\mathcal{E}^{\dagger}) =
  \begin{pmatrix}
    0.0 & 0.0&  0.0 &  0.0\\
    0.0  & 0.000932& 0.0& 0.01088\\
    0.0 & 0.0&  0.0 &  0.0\\
     0.0 &  0.01088 & 0.0 &  0.99
 \end{pmatrix},
 \label{eq:M4}
\end{equation}
where we choose $p_{24}=p_{42} > 0$ and impose $p^{n}_{24} > 0$ as a condition for solving Eq.(1).

\begin{equation}
\small
  2(\mathcal{E'}\rho_{w}(V)\mathcal{E'}^{\dagger}) =
  \begin{pmatrix}
    0.0 & 0.0&  0.0 &  0.0\\
    0.0  & 0.0& 0.0& 0.0\\
    0.0 & 0.0&  0.0105 &  0.034\\
     0.0 &  0.0 & 0.34 &  0.99
 \end{pmatrix},
 \label{eq:M4}
\end{equation}
where we choose $p_{34}=p_{43} > 0$ and impose $p^{n}_{34} > 0$ as a condition for solving Eq.(1).

\subsection{5.4 The $2(\mathcal{E'}\rho_{w}(A)\mathcal{E'}^{\dagger})$ with the $\mathcal{E'}=(\mathcal{B=A})\bigotimes\mathcal{I}$}

\begin{equation}
\small
  2(\mathcal{E'}\rho_{w}(A)\mathcal{E'}^{\dagger}) =
  \begin{pmatrix}
    0.2591 & -0.2069&  -0.2591 &  0.2069\\
    -0.2069  & 0.22934& 0.2069& -0.2293\\
    -0.2591 & 0.2069&  0.2591 &  -0.2069\\
     0.2069 &  -0.2293 & -0.2069 &  0.2293
 \end{pmatrix}.
 \label{eq:M3}
\end{equation}
 These conditions $p_{12}=p_{21} < 0$, $p_{13}=p_{31}< 0$, $p_{24}=p_{42}< 0$, and $p_{34}=p_{43}< 0$ can be used for solving  Eq.(1).

\subsection{5.5 The $\mathcal{E}$ and $\mathcal{E'}$ for $\mathcal{B}=D$}
Here we can assume $\rho_{w}(D)=\rho_{w}(A)$ because the Werner state is accidental subtracted.
\begin{equation}
\small
  2(\mathcal{E}\rho_{w}(D)\mathcal{E}^{\dagger}) =
  \begin{pmatrix}
    0.2607 & 0.2607&  0.2205 &  0.2205\\
    0.2607  & 0.2607& 0.2205& 0.2205\\
    0.2205 & 0.2205&  0.2548 &  0.2548\\
     0.2205 &  0.2205 & 0.2548 &  0.2548
 \end{pmatrix}.
 \label{eq:M4}
\end{equation}
 These conditions $p_{12}=p_{21} > 0$, $p_{13}=p_{31}> 0$, $p_{24}=p_{42}> 0$, and $p_{34}=p_{43}> 0$ can be used for solving
 Eq.(1).

\begin{equation}
\small
  2(\mathcal{E'}\rho_{w}(D)\mathcal{E'}^{\dagger}) =
  \begin{pmatrix}
    0.2392 & 0.2228&  0.2392 &  0.2228\\
    0.2228  & 0.2717& 0.2228& 0.2717\\
    0.2392 & 0.2228&  0.2392 &  0.2228\\
     0.2228 &  0.2717 & 0.2228 &  0.2717
 \end{pmatrix}.
 \label{eq:M4}
\end{equation}
 These conditions $p_{12}=p_{21} > 0$, $p_{13}=p_{31}> 0$, $p_{24}=p_{42}> 0$, and $p_{34}=p_{43}> 0$ can be used for solving Eq.(1).

\subsection{5.6 The $2(\mathcal{E'}\rho_{w}(L)\mathcal{E'}^{\dagger})$ with the $\mathcal{E'}=(\mathcal{B=L})\bigotimes\mathcal{I}$}

\begin{equation}
\fontsize{8pt}{10pt}\selectfont
  2(\mathcal{E'}\rho_{w}(L)\mathcal{E'}^{\dagger}) =
  \begin{pmatrix}
    0.2575+0.i & 0.008+i0.203&  0.0-i0.2575 &  0.2038-i0.0083\\
     0.008-i0.203  & 0.2418& -0.2038-i0.008& 0.0-i0.2418\\
  0.0 + i0.2575 &-0.2038 + i0.008&  0.2575 & 0.008+i0.2038\\
    0.2038-i0.0083 & 0.0+i0.2418 & 0.008-i0.2038 &  0.2418
 \end{pmatrix}.
 \label{eq:M3}
\end{equation}
We choose two conditions $\rm{Im}[p_{12}] > 0$ and $\rm{Im}[p_{13}] < 0$ for solving Eq.(1). Other positive values are also possible such as $\rm{Im}p_{31}]$, $\rm{Im}[p_{42}]$, and $\rm{Im}[p_{34}]$. Other negative values are also possible such as $\rm{Im}[p_{21}]$, $\rm{Im}[p_{13}]$ $\rm{Im}[p_{24}]$, and $\rm{Im}[p_{43}]$.

\subsection{5.7 The $\mathcal{E}$ and $\mathcal{E'}$ for $\mathcal{B}=R$}
Here we can assume $\rho_{w}(R)=\rho_{w}(L)$ because the Werner state is accidental subtracted.

\begin{equation}
\fontsize{8pt}{10pt}\selectfont
  2(\mathcal{E}\rho_{w}(R)\mathcal{E}^{\dagger}) =
  \begin{pmatrix}
    0.2573 & 0.0+i0.2573&  0.007-i0.2038 &  0.2038+i0.007\\
    0.0-i0.2573  & 0.2573& -0.2038-i0.007& 0.007-i0.2038\\
    0.007+i0.2038 & -0.2038+i0.007&  0.24205 &  0.0+i0.24205\\
     0.2038-i0.007 &  0.007+i0.2038 & -i0.24205 &  0.24205
 \end{pmatrix}.
 \label{eq:M4}
\end{equation}
We choose two conditions $\rm{Im}[p_{12}] > 0$ and $\rm{Im}[p_{13}] < 0$ for solving Eq.(1). Other positive values are also possible such as $\rm{Im}[p_{31}]$, $\rm{Im}[p_{42}]$, and $\rm{Im}[p_{34}]$. Other negative values are also possible such as $\rm{Im}[p_{21}]$, $\rm{Im}[p_{13}]$ $\rm{Im}[p_{24}]$, and $\rm{Im}[p_{43}]$.

\begin{equation}
\fontsize{8pt}{10pt}\selectfont
  2(\mathcal{E'}\rho_{w}(R)\mathcal{E'}^{\dagger}) =
  \begin{pmatrix}
    0.2575 & 0.0083-i0.2038&  0.0+i0.2575 &  0.2083+i0.0083\\
    0.0083+i0.2038  &0.2418& -0.2038+i0.0083& 0.0+i0.24183\\
    0.0-i0.2575 & -0.2038-i0.0083&  0.2575 &  0.0083-i0.2038\\
     0.2228 &  0.2717 & 0.0083+i0.2038 &  0.2418
 \end{pmatrix}
 \label{eq:M4}
\end{equation},
where we choose two conditions $\rm{Im}[p_{12}]< 0$ and $\rm{Im}[p_{13}] > 0$ for solving Eq.(1). Other negative values also possible such as $\rm{Im}[p_{31}]$, $\rm{Im}[p_{42}]$, and $\rm{Im}[p_{34}]$. Other positive values are also possible such as $\rm{Im}[p_{21}]$, $\rm{Im}[p_{13}]$ $\rm{Im}[p_{24}]$, and $\rm{Im}[p_{43}]$.

\subsection{5.8 Mathematica Function $\texttt{NSolve}$}
For the detection of classical state $H$,  we use the $\texttt{NSolve}$ as given by,\par
NSolve[$p^{n}_{11}$==0.416 \&\& $p^{n}_{14}$==0.282 \&\& $p^{n}_{23}$==-0.00785 \&\& $p^{n}_{13}>0$ \&\& $1>x>0$, $\{x,c,a\}$].

For the detection of classical state $A$,  we use the $\texttt{NSolve}$ as given by,\par
NSolve[$p^{n}_{11}$==0.349 \&\& $p^{n}_{14}$==0.206 \&\& $p^{n}_{23}$==0.0219 \&\& $p^{n}_{\rm{off\,diag.}} < 0$ \&\& $1>x>0$,$\{x,c,a\}$].

For the detection of classical state $L$,  we use the $\texttt{NSolve}$ as given by,\par
NSolve[$p^{n}_{11}$==0.275 \&\& $p^{n}_{14}$==0.0535 \&\& $p^{n}_{23}$==-0.00110 \&\& $\rm{Im}[p^{n}_{12}] < 0$ \&\& $\rm{Im}[p^{n}_{13}] > 0$ \&\& $1>x>0$,$\{x,c,a\}$].

\section{Data availability}
The data that supports the findings of this study are available from the corresponding author upon reasonable request.

\newcommand{\noopsort}[1]{} \newcommand{\printfirst}[2]{#1}
   \newcommand{\singleletter}[1]{#1} \newcommand{\switchargs}[2]{#2#1}

\section{Acknowledgments}

This work was supported in part by the DOE (DE-SC0020537)

\section{Author Contributions}
K.F.L and P.K conceived the research.  K.F.L analyzed the data, wrote the paper and prepared the manuscript.

\section{Competing Interests}
The authors declare no competing interests.

\end{document}